\documentclass[fleqn,12pt,twoside]{article}
\usepackage{espcrc1}

\usepackage{graphicx}
\usepackage[figuresright]{rotating}

\newcommand{\AmS}{{\protect\the\textfont2
  A\kern-.1667em\lower.5ex\hbox{M}\kern-.125emS}}
\newcommand{\sqrtsNN}{\mbox{$\sqrt{\mathrm{s}_{_{\mathrm{NN}}}}$}}

\title{Identical Particle Interferometry at STAR}

\author{M. L\'{o}pez Noriega\address{The Ohio State University, 174 W 18th Ave., Columbus, Ohio 43210, USA}
for the STAR Collaboration\footnote{For the full author list and
acknowledgements, see Appendix "Collaborations" of this volume}}

\begin{document}

\maketitle

\begin{abstract}
We present preliminary results from a two-particle intensity
interferometry analysis of charged pions and neutral kaons emitted
from Au+Au collisions at {\sqrtsNN}~= 130 and 200 GeV measured in
the STAR detector at RHIC. The dependence of the apparent pion
source on beam energy, multiplicity, transverse momentum and
emission angle with respect to the reaction plane are discussed.
\end{abstract}

\section{Introduction}
Two-particle intensity interferometry (HBT) is a useful tool to
study the space-time geometry of the particle-emitting source in
heavy ion collisions \cite{BAUER92,HEINZ99}. It also contains
dynamical information that can be explored by studying the
transverse momentum ($p_{T}$) dependence of the apparent source
size \cite{PRATT84,MAKHL88}. In non-central collisions information
on the anisotropic shape of the pion-emitting region can be
extracted by measuring two-pion correlation functions as a
function of emission angle with respect the reaction plane
\cite{WIEDE98}.

In this paper we present two-pion correlation systematics as a
function of the collision energy {\sqrtsNN}, transverse mass
($m_{T} = \sqrt{p_{T}^{2} + m^{2}}$), multiplicity, and emission
angle with respect to the reaction plane in Au+Au collisions
produced by the Relativistic Heavy Ion Collider (RHIC) at
Brookhaven National Laboratory. We also present the first
significant HBT measurement for neutral kaons from heavy ion
collisions; besides carrying important information about the
dynamics of strange particles, these measurements provide valuable
cross-checks on charged kaon HBT, as complication due to various
corrections (e.g. two-track efficiency and Coulomb correction) are
quite different for neutral particles.

Experimentally, two-particle correlations are studied by
constructing the correlation function
$C_{2}$(\textbf{q})~=~A(\textbf{q})/B(\textbf{q}). Here
A(\textbf{q}) is the measured distribution of the momentum
difference \textbf{q}~=~$\textbf{p}_{1}$~-~$\textbf{p}_{2}$ for
pairs of particles from the same event, and B(\textbf{q}) is the
corresponding distribution for pairs of particles from different
events.

\section{Experimental Details}
For this analysis we selected events with a collision vertex
position within $\pm$30~cm measured from the center of the 4~m
long STAR Time Projection Chamber (TPC), and we mixed events only
if their longitudinal primary vertex positions were no farther
apart than 6~cm. We divided our sample into three centrality bins,
where the centrality was characterized according to the measured
multiplicity of charged particles at midrapidity. The three
centrality bins correspond to 0-10$\%$ (central), 10-30$\%$
(midcentral), and 30-70$\%$ (peripheral) of the total hadronic
cross section. Charged pions were identified by correlating their
specific ionization in the gas of the TPC with their measured
momentum. Neutral kaons were identified via topological methods
\cite{ADLER02}.

For charged pions, the effects of track-splitting (reconstruction
of a single track as two tracks) and track-merging (two tracks
with similar momenta reconstructed as a single track) were
eliminated as described in \cite{ADLER01}. We applied to each
background pair a Coulomb correction \cite{PRATT90} corresponding
to a spherical Gaussian source of 5~fm radius.

In the neutral particle analysis, the possibility of a single
$K^{0}_{s}$ being correlated with itself was eliminated by
requiring that a pair of $K^{0}_{s}$'s have unique daughters, and
that their decay positions were spatially well-separated.

\section{Pion HBT versus $p_{T}$ and centrality}

\begin{figure}
\begin{minipage}[t]{0.55\textwidth}
\includegraphics[width=\textwidth,height=0.90\textwidth]{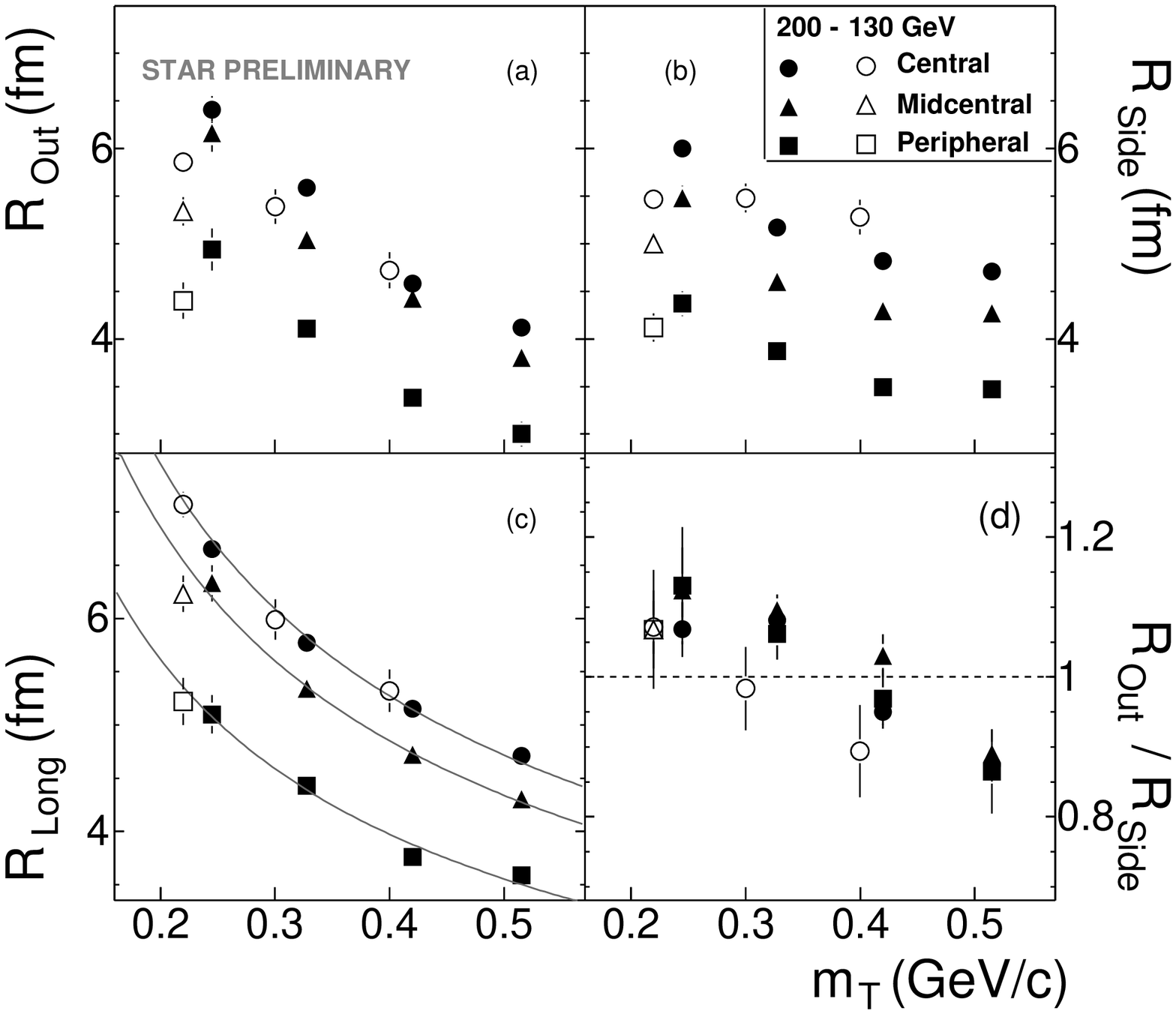}
\caption{Centrality dependence of the $m_{T}$ dependence of the
$\pi^{-}\pi^{-}$ HBT radii for {\sqrtsNN}=200 GeV in closed
symbols. The error bars indicate the statistical errors only.
Multiplicity and $m_{T}$ dependence of the HBT radii of the
$\pi^{-}\pi^{-}$ HBT radii for {\sqrtsNN}=130 GeV in open symbols.
Lines in (c) represent the Mahklin/Sinyukov fit \cite{MAKHL88}.}
\label{HBTparam}
\end{minipage}\hfill
\begin{minipage}[t]{0.41\textwidth}
\includegraphics[width=\textwidth,height=1.15\textwidth]{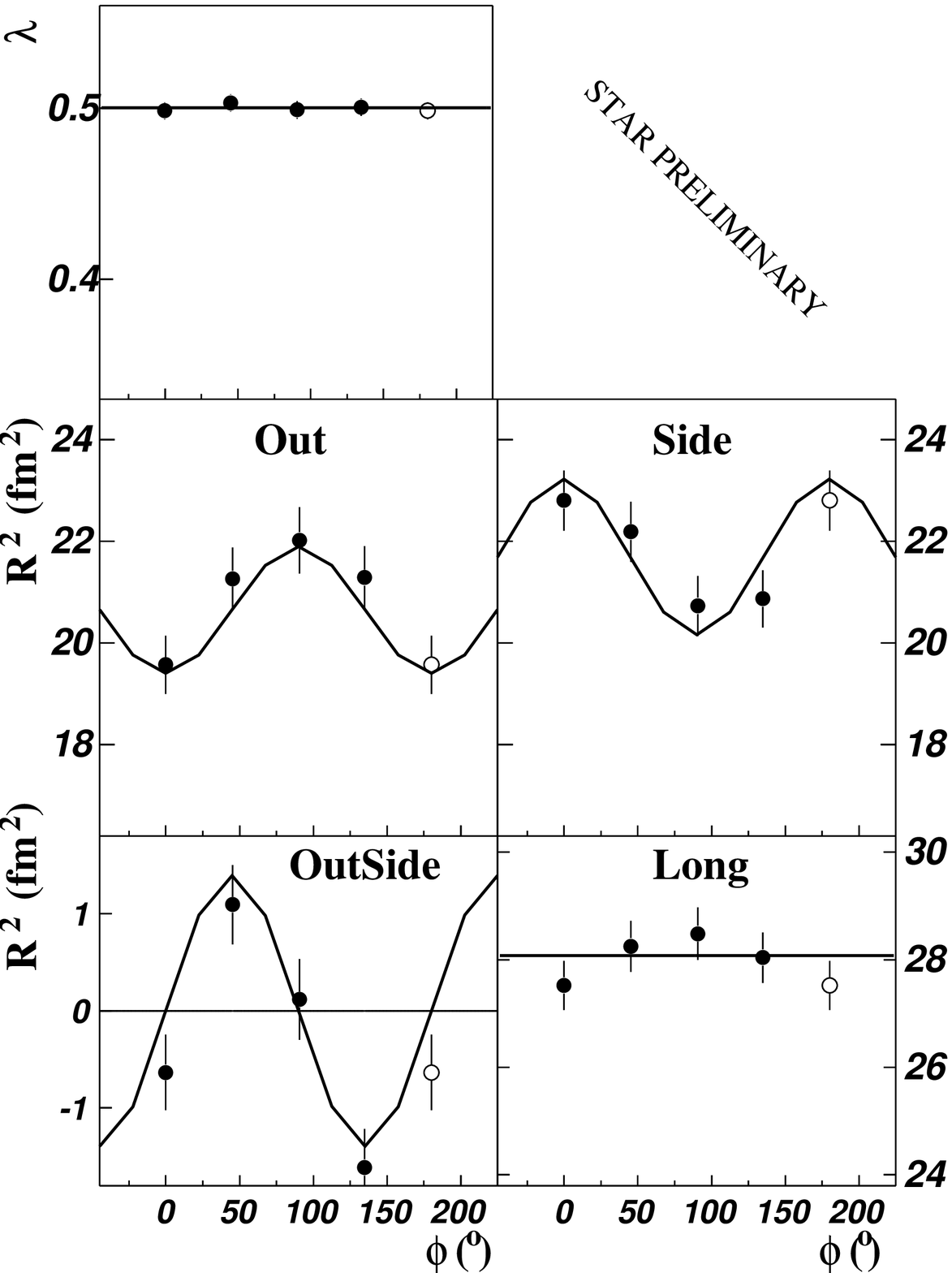}
\caption{Azimuthal dependence of HBT parameters from peripheral
Au+Au collisions at {\sqrtsNN}=130 GeV. Bolstered statistics by
summing results of $\pi^{-}$ and $\pi^{+}$ analysis. Lines
represent the blast wave source parameterization.} \label{RPHBT}
\end{minipage}\hfill
\end{figure}

The three-dimensional correlation functions were generated. The
relative momentum \textbf{q} was measured in the longitudinal
co-moving system (LCMS) frame, and decomposed according to the
Pratt-Bertsch \cite{PRATT90,BERTS89} "out-side-long"
parametrization. The correlation functions were fit with:
C($q_{o}$,~$q_{s}$,~$q_{l}$)~=~1~+
$\lambda$~$\exp$(-$R_{o}^{2}$~$q_{o}^{2}$~-~$R_{s}^{2}$~$q_{s}^{2}$~-~$R_{l}^{2}$~$q_{l}^{2}$).

The effect of the single-particle momentum resolution ($\delta$p/p
$\sim$ 1$\%$ for the particles under study) induces systematic
underestimation of the HBT parameters. Using an iterative
procedure \cite{ADLER01}, we corrected our correlation functions
for finite resolution effects. The correction due to the
uncertainty on the removal of the artificial reduction of the HBT
parameters associated with the anti-merging cut is still being
finalized; we estimate systematic errors on HBT radii at about
1~fm.

Fig.\ref{HBTparam} shows the $m_{T}$ dependence of the source
parameters for negative pions at three centrality bins. The three
radii increase with increasing centrality and $R_{l}$ varies
similarly to $R_{o}$, $R_{s}$; for $R_{o}$ and $R_{s}$ this
increase may be attributed to the geometrical overlap of the two
nuclei. The extracted radii rapidly decrease as a function of
$m_{T}$, which is an indication of transverse flow \cite{TOMAS00};
interestingly the $m_{T}$ dependence is independent of centrality.

While slightly larger at 200 GeV, the transverse homogeneity
lengths are similar for both energies (Figs.\ref{HBTparam} (a) and
(b)). These results suggest a steeper fall-off with $m_{T}$, which
might indicate increased flow; finalized corrections are necessary
before drawing such conclusions. $R_{o}$ falls somewhat steeper
than $R_{s}$ with $m_{T}$ and $R_{o}/R_{s}\sim1$
(Fig.~\ref{HBTparam}(d)) which indicates short emission duration,
$\Delta$t $\sim$ 1-2 fm/c, in a blast wave fit \cite{RETIE01}.

The longitudinal radius falls along the same curve at both
energies (Fig.~\ref{HBTparam}(c)). Assuming boost-invariant
longitudinal flow, we can extract an evolution timescale, by using
a simple fit \cite{MAKHL88} (solid lines Fig.~\ref{HBTparam}(c)):
$R_{l}= <t_{fo}>\sqrt{\frac{T_{k}}{m_{T}}}$. Taking $T_{k}$ = 110
MeV we got $<t_{fo}>$~$\approx$~10 fm/c for central events and
$<t_{fo}>$~$\approx$~7.6 fm/c for peripheral events. Hence, the
evolution time, in addition to the emission duration, is quite
short.

\begin{figure}
\centering
\begin{minipage}[c]{0.45\textwidth}
\centering
\includegraphics[width=\textwidth,height=0.55\textwidth]{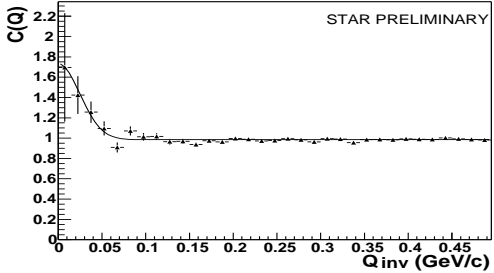}
\end{minipage}\hfill
\begin{minipage}[c]{0.53\textwidth}
\centering \caption{One-dimensional correlation function
C($Q_{inv}$) for $K^{0}_{s}$ candidates with invariant mass within
$\sim\pm$ 15 MeV/c$^{2}$, $0.1<p_{T}<3$ GeV/c at midrapidity from
central Au+Au collisions at {\sqrtsNN}=200 GeV. Line represents
the fit to a one-dimensional Gaussian.} \label{K0CF}
\end{minipage}
\end{figure}

\section{Pion HBT with respect the reaction plane}

Azimuthally-sensitive HBT, correlated event-by-event with the
reaction plane, provides information about the coordinate-space
anisotropies of the emitting source \cite{TEANE99}.

We calculated the three-dimensional correlation functions with
fixed pair angle with respect to the $2^{nd}$-order event plane
(determined from elliptic flow)
$\phi$~=~$\angle(\textbf{K}_{\bot},\textbf{b})$, where
$\textbf{K}_{\bot}$~=~$(\textbf{p}_{1}+\textbf{p}_{2})_{\bot}$ is
the total momentum of the pair perpendicular to the beam. The
correlation functions for each $\phi$ bin were fit with the
standard Gaussian parametrization
C(\textbf{q},$\phi$)~=~1~+~$\lambda(\phi)\exp[-q_{i}q_{j}R_{ij}^{2}(\phi)]$,
in this case the cross-term $R_{os}^{2}$ is also relevant.

Fig.\ref{RPHBT} shows the HBT parameters plotted as a function of
$\phi$ at {\sqrtsNN}=130 GeV. Data were corrected for event plane
resolution and merging systematics. $R_{o}$ and $R_{s}$ show
significant equal and opposite $2^{nd}$-order oscillations with
phases indicating an out of plane extended source geometry. This
would suggest a fast evolution, since positive $v_2$ will push the
source from an out-of-plane extended configuration (present in the
entrance channel and caused by the partial overlap of the
colliding nuclei) to in-plane extended. For comparison, the lines
represent a fit with the "blast wave model" \cite{RETIE01},
corresponding to a transversely deformed (5$\%$ extended
out-of-plane) source geometry with an oscillating flow field, and
a short (2 fm/c) emission duration. This model successfully
reproduces the details of elliptic flow measured by STAR
\cite{ADLER01B}.

Preliminary results for $\pi^{-}$ and $\pi^{+}$ from peripheral
Au+Au collisions at {\sqrtsNN}=200 GeV confirm the oscillations
measured at 130 GeV and the higher statistics of these
measurements will allow $m_{T}$ and centrality systematic
analysis.

\section{$K^{0}_{s}$ Interferometry}

The good efficiency of the TPC for finding and reconstructing
short-lived neutral kaons gives us the opportunity to study HBT
correlations for $K^{0}_{s}$. This analysis should allow us to
extend our systematics to higher $p_{T}$. Using the topology of
the decay $K^{0}_{s} \rightarrow \pi^{+}\pi^{-}$ we reconstructed
$\sim$3.8 $K^{0}_{s}$ candidates per event with $<m_{T}> \sim$
1.12 GeV/c$^{2}$. Fig.\ref{K0CF} shows the one-dimensional
correlation function which exhibits a promising low-Q correlation
signal. Fitting with the functional form C(Q) = 1 +
$\lambda\exp(-Q^{2}R^{2})$ yields $\lambda = 0.76 \pm 0.29$ and $R
= 5.75 \pm 1$ fm. If these results persist as the analysis is
finalized, it is interesting to speculate on possible causes of
such a large homogeneity length at this high $m_{T}$.

\section{Conclusion}

We have presented identical meson interferometry results for Au+Au
collisions at {\sqrtsNN}=200 GeV. With respect to multiplicity and
$m_{T}$ dependencies, pion HBT radii are very similar to results
reported at {\sqrtsNN}=130 GeV. Our results indicate that both the
evolution timescale (as measured by the $m_{T}$ dependence of
$R_{l}$ and the out-of-plane extended source indicated by the
azimuthally-sensitive HBT) and the emission duration (probed by
comparing $R_{o}$ to $R_{s}$) are surprisingly fast. The large
acceptance and excellent tracking of the TPC has allowed for the
first solid measurement of $K^{0}_{s}$-$K^{0}_{s}$ correlations.
While still preliminary, these suggest a surprisingly large
homogeneity length at high $m_{T}$.

\end{document}